\documentclass[conference]{IEEEtran}
\ifCLASSINFOpdf
\else
\fi
\usepackage{graphicx}
\usepackage{setspace}
\usepackage{amsmath}
\usepackage{amssymb} 
\usepackage{bm} 
\usepackage{cite}
\usepackage{float}
\usepackage[usenames,dvipsnames]{pstricks}
\usepackage{epsfig}
\usepackage{pst-grad} 
\usepackage{pst-plot} 
\usepackage{tabularx}
\usepackage[keeplastbox]{flushend}
\usepackage[T1]{fontenc}
\newcommand{\lb} {\left}
\newcommand{\rb} {\right}

\allowdisplaybreaks
\flushend
\usepackage{caption}
\usepackage{subcaption}
\usepackage[utf8]{inputenc}

\begin{document}
\onecolumn{\noindent © 2020 IEEE. Personal use of this material is permitted. Permission from IEEE must be obtained for all other uses, in any current or future media, including reprinting/republishing this material for advertising or promotional purposes, creating new collective works, for resale or redistribution to servers or lists, or reuse of any copyrighted component of this work in other works. \\

\noindent Conference: IEEE Global Communications Conference 2020, Taipei, Taiwan , 7-11 December 2020 \\

\noindent DOI: 10.1109/GLOBECOM42002.2020.9322432 \\

\noindent URL: https://ieeexplore.ieee.org/document/9322432\\
}

\twocolumn{
\title{Recurrent Neural Network Assisted Transmitter Selection for Secrecy in Cognitive Radio Network}


\author{
    \IEEEauthorblockN{Shalini Tripathi\IEEEauthorrefmark{1}, Chinmoy~Kundu\IEEEauthorrefmark{2}, 
    Octavia A. Dobre\IEEEauthorrefmark{3},
     Ankur Bansal\IEEEauthorrefmark{4}, and
     Mark F. Flanagan\IEEEauthorrefmark{5}}
     
    \IEEEauthorblockA{\IEEEauthorrefmark{1}\IEEEauthorrefmark{4}Department of EE, Indian Institute of Technology Jammu, India }
    
     \IEEEauthorblockA{\IEEEauthorrefmark{2}\IEEEauthorrefmark{5}School of Electrical and Electronic Engineering, University College Dublin, Ireland}
         
  \IEEEauthorblockA{\IEEEauthorrefmark{3}Engineering and Applied Science, Memorial University, Canada}
           
    \textrm{\{\IEEEauthorrefmark{1}2019ree0001,\IEEEauthorrefmark{4}ankur.bansal\}@iitjammu.ac.in},
    {\IEEEauthorrefmark{2}chinmoy.kundu@ucd.ie}, {\IEEEauthorrefmark{3}odobre@mun.ca}, {\IEEEauthorrefmark{5}mark.flanagan@ieee.org}}

\maketitle



\begin{abstract}
In this paper, we apply the \emph{long short-term memory} (LSTM), an advanced recurrent neural network based machine learning (ML) technique, to the problem of transmitter selection (TS) for secrecy in an underlay small-cell cognitive radio network with unreliable backhaul connections. The cognitive communication scenario under consideration has a secondary small-cell network that shares the same spectrum of the primary network with an agreement to always maintain a desired outage probability constraint in the primary network. Due to the interference from the secondary transmitter common to all primary transmissions, the secrecy rates for the different transmitters are correlated. 
LSTM exploits this correlation and matches the performance of the conventional technique when the number of transmitters is small. As the number grows, the performance degrades in the same manner as other ML techniques such as support vector machine, $k$-nearest neighbors, naive Bayes, and deep neural network. However, LSTM still significantly outperforms these techniques  in misclassification ratio and secrecy outage probability. It also reduces the feedback overhead against conventional TS.






\end{abstract}

\section{Introduction}
The pioneering work of Wyner in information-theoretic security has led to the current popularity of wireless physical layer security \cite{wyner_wiretap}. This has shown promising results in reducing complexity against traditional key-based cryptographic
techniques without requiring
secret key sharing, management, and complex algorithms. Enhancing security of a communication network by improving  selection diversity is a simple alternative against beamforming, jamming or noise forwarding. This is extensively applied in transmit antenna selection (TAS) in multiantenna systems, transmitter selection (TS) in multiple source systems, and relay selection (RS) in cooperative systems \cite{Yang_Schober_TAS,  Kundu2016Relay, Kundu_relsel}. Next-generation wireless networks will be highly dense and heterogeneous due to the increasing demand for high data rate applications. The combination of spectrum sharing in cognitive radio (CR) and small-cell networks with wireless backhaul is one of the  probable solutions. Although wireless backhaul may be a cost-effective alternative, it suffers from reliability issues.  The security of such a network is of paramount importance. A comprehensive analysis of secrecy enhancement through TS in a cognitive small-cell network with uncertain wireless backhaul can be found in \cite{Vu2017Secure, Kundu_TVT19}.  

With the increasing application of machine learning (ML) in many different domains, the application interest in the area of TAS problems in wireless communications is also accelerating  \cite{Joung2016,LiuTAS_MIMO,  Yao_ML, Wang_DT}. 
The author in \cite{Joung2016} first proposed to solve the TAS problem as a multiclass-classification
learning problem using ML. Through the application of $k$-nearest neighbors ($k$-NN) and support
vector machine (SVM) algorithms, the author showed that the computational complexity and feedback overhead can be reduced compared to the traditional wireless techniques, while maintaining a reasonable performance accuracy.  
In \cite{LiuTAS_MIMO}, the authors extended the idea of solving the TAS problem through multiclass-classification ML algorithms in the multiple-input multiple-output (MIMO) wiretap channel using SVM and naive Bayes (NB) schemes. These three ML algorithms were also applied to TAS in an untrusted relay networks in \cite{Yao_ML}.  Recently, in \cite{Wang_DT}, the multiclass-classification approach was also extended to RS problems in dual-hop wireless networks
using a decision-tree-based
ML scheme. We refer to these algorithms as ``traditional'' ML algorithm as they require explicit programming to extract features.

Deep learning (DL), a subclass of ML, can also be used to predict and classify from complex raw data without being explicitly programmed to extract features as in ML \cite{Zhang}. 
DL can handle nonlinear problems and its performance improves with the size of the data, which may not be the case for traditional ML models \cite{Goodfellow-et-al-2016}. The authors in \cite{Yao_DNN}  implemented a deep neural network (DNN) scheme for TAS in an untrusted relay network. The authors showed that the DNN performs better than traditional ML schemes and achieves almost the same secrecy rate as the conventional scheme. In \cite{Ibrahim_AS_DNN}, the authors proposed a DNN-based approach to implement the joint multicast beamforming and TAS problem 
to reduce the computational complexity. An artificial neural network structure along with a decision-tree-based approach was proposed for TAS in \cite{Gecgel_DL-MLP_TAS} to improve the
error performance in the presence of time-correlated channels
and channel estimation errors.

The drawback of DNN is that it has no memory and therefore cannot retain information regarding previous computations. As a result, DNN does not work well with dependencies in data set. Conventional recurrent neural networks (RNNs) can overcome this challenge by having recurrent connections between previous and current computations \cite{Goodfellow-et-al-2016}. An RNN has short-term memory and cannot handle long-term dependencies in the data set due to vanishing gradient and exploding gradient problems \cite{Hochreiter}. Long short-term memory (LSTM), as an advanced version of  conventional RNN, can outperform them in cases where the data set has long-term dependencies  \cite{Hochreiter, Greff_LSTM}. 
Thus, LSTM is widely applied in large-scale acoustic modeling, speech recognition, and text categorization \cite{Sak_RNN, Graves}.
In wireless communications, LSTM is applied for the channel state information (CSI) prediction for 5G wireless communication in  \cite{LuoCSI_LSTM_CNN} and for signal detection in multipath MIMO environment \cite{BaekCSI_LSTM}, where LSTM based detection performed best.   
Despite its suitability in capturing long-term dependencies in the data set, LSTM has not yet been considered  for the TAS/TS/RS problem through multiclass-classification. 

In this paper, we propose LSTM as a tool for solving the TS problem to improve the secrecy of a CR network with wireless backhaul.
We assume that interference exists from both primary to secondary network and vice versa.  The channel fading model is considered to be independent non-identically distributed (INID) complex Gaussian, in contrast to the channel assumptions considered in \cite{Vu2017Secure, Kundu_TVT19}. Due to the common interference term from the primary network to the secondary receiver, the secrecy rates of the individual secondary transmitters are correlated. 
We propose 
LSTM for our TS problem to exploit the long-term dependency generated in data due to this interference. 
As the number of transmitters increases, the performance of all the learning models, including LSTM, degrades compared to the conventional search. This has not been pointed out by the earlier literature in ML multiclass-classification problems. We will show that due to its ability to capture long-term data dependency, LSTM has a significantly lower misclassification ratio than other learning models and consequently superior secrecy outage probability (SOP). LSTM can also reduce the feedback overhead by half compared to the conventional scheme.

The rest of the paper is organised as follows. In Section II, the system model is described. The SOP for the optimal selection is defined in Section III. The training methodology for the ML based transmitter selection is described in Section IV. Sections V and VI provide computational complexity and numerical results, respectively. Finally, the paper is concluded in Section VII. 

\textit{Notation:} $\mathbb{P}[\cdot]$ is the probability of occurrence of an event. 
 $\mathbb{E}[\cdot]$ denotes expectation operation.
The channel coefficient between any two nodes $A$ and $B$ is denoted by $h_{AB}$ and  the signal-to-interference-plus-noise-ratio (SINR) at $B$ for the link $A$-$B$ is denoted as $\Gamma_{AB}$.


\section{System Model}
\begin{figure}
 \centering 
 \includegraphics[width=2.2in]{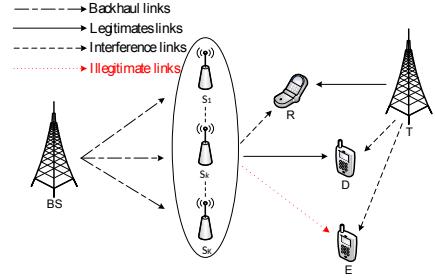} 
 \caption{Cognitive radio network with unreliable backhaul.}
 \vspace{-.7cm}
 \label{fig_system}
 \end{figure}
The system model, depicted in Fig. \ref{fig_system}, consists of a secondary small-cell network and a primary cellular network where the former is sharing the spectrum of the latter using underlay CR technology. In the secondary network, a macro-cell base station (BS) provides wireless backhaul connections to $K$ small-cell transmitters which in turn serve a destination $D$. A passive eavesdropper $E$ is present in the secondary network listening to the communication from secondary transmitters. The primary network consists of a transmitter $T$ and its receiver $R$. As both the primary and secondary networks share the same spectrum, interference from the primary to secondary network, as well as from the secondary to primary network, is considered. The outage probability of the primary network is considered as the QoS constraint.
The channel coefficients of the links, $S_k$-$D$,  $S_k$-$E$ and $S_k$-$R$, for each $k$, and $T$-$R$, $T$-$D$, and $T$-$E$, have INID complex Gaussian  distribution with zero mean and variance $1/\lambda_{s_kd}$, $1/\lambda_{s_ke}$,  $1/\lambda_{s_kr}$,  and $1/\lambda_{tr}$, $1/\lambda_{td}$, and $1/\lambda_{te}$, respectively. Noise at all receivers is considered as complex additive white Gaussian noise (AWGN) with equal parameters of zero mean and variance $N_0$.         

The SINR at the secondary destination or eavesdropper due to the interference from the primary transmitter can be expressed as
\begin{align}\label{SNR_E}
&\Gamma_{SB}=\dfrac{P_S  |h_{S_{k^{\ast}}B}|^2}{P_T  |h_{TB}|^2+N_0}\, ,
\end{align} 
where $P_S$ is the maximum transmit power allowed for the small-cell transmitter satisfying the primary QoS constraint, $P_T$ is the primary transmit power at $T$, $k^{\ast}$ signifies the optimal secondary transmitter, and $B\in\{D, E\}$.

\subsection{Modelling Wireless Backhaul Uncertainty} The backhaul links  between the BS and the small-cell transmitters are unreliable due to the wireless propagation channel, and thus, have a non-zero probability of failure. The backhaul uncertainty is modeled by INID Bernoulli random variables $\mathbb{I}_k$ for $k=1,\ldots, K$, with success probability $\mathbb{P}(\mathbb{I}_k=1)=\delta_k$ and failure probability  $\mathbb{P}(\mathbb{I}_k=0)=1-\delta_k$  $ \forall k=1\ldots, K$. 
In contrast to \cite{Kundu_TVT19}, the optimal transmitter is selected with the knowledge of active backhaul links. This implies that the optimal transmitter lies within the set of transmitters whose backhaul links are active at a certain point in time.   

\subsection{Secondary Transmit Power Constraint}
Due to the CR approach considered for simultaneous primary and secondary transmission, the secondary transmission power is restricted to guarantee a certain primary QoS. The primary constraint is taken to be the outage probability and should be below a threshold level. The secondary transmit power constraint can be obtained from the outage probability constraint of the primary receiver as
\begin{align}\label{desired outage probability}
\mathbb{P}\left[{\Gamma_{TR}}  < \Gamma_0\right]\leq \Phi,
\end{align}
where 
$0 < \Phi <1$ is the primary outage probability constraint, 
$\Gamma_0=2^{R_{th}}-1$ is the SINR threshold for the primary outage,  with $\beta$ as the threshold rate of the primary outage, and $\Gamma_{TR}$ is given as \begin{align}\label{SINR at P_R}
\Gamma_{TR}= \frac{P_T|h_{TR}|^2}{P_S|h_{S_{k^*}R}|^2+N_0}.
\end{align}
As such, $P_S$ can be evaluated from (\ref{desired outage probability}) using the cumulative distribution function of $\Gamma_{TR}$ and is given in \cite{Kundu_TVT19}.


\section{SOP for Optimal Selection}
In this section we present the performance metric for the optimal TS.
The optimal transmitter is defined as the one which provides maximum instantaneous secrecy rate. Considering the backhaul uncertainty, and assuming backhaul activity knowledge is available before selection, the optimal transmitter is selected as 
\begin{align}\label{OS_scheme}
k^* = \arg \max_{k} \{ {\mathbb{I}_k C^{k}_s}\}=\arg \max_{k\in\mathcal S} \{ C^{k}_s\},
\end{align} 
where $\mathcal S$ is the set of transmitters with active backhaul links, and $C^{k}_s$ is the secrecy rate of the wiretap channel formed by the pair of links $S_k$-$D$ and $S_k$-$E$, which is expressed as 
  \begin{align}
  \label{eq_max_snr_ratio}
  C^{k}_s=\max\lb\{\log_2\lb[\frac{1+\Gamma_{S_kD}}{1+\Gamma_{S_kE}}\rb], 0\rb\} .
  \end{align}
Equation (\ref{OS_scheme}) states that if backhaul knowledge is available, then optimal selection can be performed within the set of transmitters with active backhauls. To perform optimal TS using (\ref{OS_scheme}), global CSI is required. Both real and imaginary parts of the complex channel coefficients are required to be fed back to a central decision making unit. 

To derive the performance of the optimal selection scheme, we will find the SOP. This is defined as the probability that the maximum achievable secrecy rate through selection is below a predefined threshold $R_{th}$. 
Including backhaul reliability knowledge, the SOP can be evaluated as 
 \begin{align}
 \label{OP_OS_WITH}
 \mathcal{P}_{out}(R_{th})&=\sum\limits_{\mathcal{S} \subseteq \{ 1,2,\ldots,K \}}  \mathbb{P} [\mathcal S] \mathbb P\left[ \max_{k \in \mathcal S}\{C_{s}^k\} < R_{th}\right],
 \end{align}
where the probability that the active backhaul set is $\mathcal{S}$ is given by 
\begin{align}
\label{eq_prob_active_set}
  \mathbb{P} [\mathcal S] = \prod_{i\in \mathcal S } \delta_i \prod_{j \notin {\mathcal S}} (1-\delta_j).
  \end{align}
In (\ref{eq_prob_active_set}), ${\mathcal S}$ is the inactive backhaul set. 
Note that the evaluation of (\ref{OP_OS_WITH})  requires the computations of the summation which involves $2^K$ terms. 
Also, it is difficult to derive a closed-form expression of $\mathbb P\left[ \max_{k \in \mathcal S}\{C_{s}^k\} < R_{th}\right]$ in (\ref{OP_OS_WITH}) due to two reasons. The first one is that the links are not identically distributed, and the second is that the values $C_s^k$ for different $k$ are correlated due to the involvement of the common terms $h_{TD}$ and $h_{TE}$. Therefore, in this paper we focus on developing various ML based techniques for optimal selection and use SOP as a performance measure.

\section{ML Based Transmitter Selection}
We utilize ML data classification property to design a TS mechanism which maximizes the instantaneous secrecy capacity in (\ref{OS_scheme}). 
We shall construct a training data set using the relevant channel gains that affect the instantaneous secrecy capacity and will feed to an ML based classifier to design a classification model. Once the model is developed, if we provide a random sample input, the ML based classifier should predict the transmitter that maximizes the instantaneous secrecy capacity.
The process of data set generation, labeling, and prediction  are explained next.

\subsection{Construction of the Training Data Sets} 
To construct the training data set, a three-step procedure is followed as detailed below.

    {\it Data set generation}: In this step, we generate the input variables for training a learning model. From (\ref{OS_scheme}) and (\ref{eq_max_snr_ratio}), we observe that TS depends on $h_{S_{k}D}$, $h_{S_{k}E}$, $h_{TE}$, and $h_{TD}$, where $k\in\{1, \ldots K\}$ for a given $\Phi$, $R_{th}$, and $P_T$. Therefore, we design $M$  training data set matrices of dimension $4\times K$, $\mathbf{D}^m$, where $m\in\{1,\ldots,M\}$, such that its columns represent CSI data values corresponding to each $k$, i.e., $[h_{S_{k}D}, h_{S_{k}E}, h_{TE}, h_{TD}]^\mathbb{T}$ for $k\in{1,\cdots,K}$. Here, $\mathbb{T}$ denotes the transpose operator. As traditional ML models and the DNN model take a vector as input for the learning process, we design a vector $\mathbf{d}^m$ of length $N=4K$ by arranging the columns of $\mathbf{D}^m$ one after the other. $\mathbf{d}^m$ is called the feature vector. As learning models act on real-valued data, it is necessary to manipulate the complex CSI values to obtain real-valued data. In addition, training data should be normalized to avoid significant learning bias. This process is divided into the following steps:
    
    \begin{enumerate}
        
        \item We find the absolute values of each complex CSI element to generate learning compatible real-valued feature vector,  $\mathbf{d}^m$, $m \in \{1, \cdots, M\}$ from earlier $\mathbf{d}^m$. As learning models require only absolute values of the complex channel coefficients, we can reduce the feedback overhead by half compared to feeding back complex coefficients.  
        
        \item We normalize the real-valued feature vector $\mathbf{d}^m$ via
        \begin{align}
            t_i^m =  \frac{d_i^m - \mathbb{E}[\mathbf{d}^m]}{\max(\mathbf{d}^m)-\min(\mathbf{d}^m)},
        \end{align}
        where $t_i^m$ and $ d_i^m $ are the $i$th element of $ \mathbf{t}^m$ and $ \mathbf{d}^m$, respectively, $i \in \{1, \ldots, 4K\}$, $ \mathbf{t}^m$ is the normalized feature vector. From $\mathbf{t}^m$ we obtain the normalized feature matrix $\mathbf{T}^m$ following the reverse process of generating $\mathbf{D}^m$ to $\mathbf{d}^m$.    
        
    \end{enumerate}
    
    {\it Key performance indicator (KPI) design}: A KPI is intended for labelling training samples. In this paper, maximizing the instantaneous secrecy capacity of the system with backhaul uncertainty is the goal; hence, our KPI is the term $(\mathbb{I}_k C_s^k)$ in (\ref{OS_scheme}).
    
    {\it Labeling}: We seek to find the maximum secrecy capacity among $K$ transmitters considering backhaul uncertainty; therefore, we decide there should be $\ell=(K+1)$ possible labels. The label should indicate the index of the transmitter with maximum secrecy capacity only when its backhaul is also active. The $(K+1)$th label is kept for the case when no active transmitters are available due to backhaul inactivity. We calculate the $K$ KPIs using the feature vector $\mathbf{d}^m$ and label each vector with one of the $(K+1)$ possible labels. A label vector is generated having $M$ labels for each $\mathbf{d}^m$ where $m\in \{1,\ldots,M \}$. 

\subsection{Model Prediction}
We train the traditional ML and DNN with $m$ data sets $\mathbf{t}^m$; in contrast, LSTM is trained with $\mathbf{T}^m$ along with the corresponding $M$ labels. Each column of $\mathbf{T}^m$ is correlated due to the common interference terms $h_{TD}$ and $h_{TE}$. This introduces long-term dependency in the data set if $K$ is large. The trained network is then used for label prediction with the normalized test feature vector or matrix. The output of the model provides the predicted label corresponding to the index of the selected transmitter. 



\section{Computational Complexity}
In this section, the prediction complexity of each learning model is analyzed as the training complexity can be trained offline. The complexities for SVM, $k$-NN, NB, and DNN are presented in the Table \ref{table1} following  \cite{Joung2016, LiuTAS_MIMO}, and  \cite{Yao_ML}, respectively. 
In the DNN scheme, $L_1 = 256$ and $L_2=128$ are the number of neurons in the first and second hidden layers, respectively \cite{Yao_ML}. For the standard LSTM network, 
the computational complexity of each time step
(in our case per transmitter due to our construction of the feature matrix, $\mathbf{D}
^m$) is $\mathcal{O}(W)$, where $\mathcal{O}(\cdot)$ is defined as the ``order of'' symbol. Since the number of transmitters is $K$, the complexity of our proposed  LSTM model is $\mathcal{O}(KW)$. Here $W$ represents the total number of parameter computations assuming one memory cell per memory block while ignoring the bias neuron \cite{Sak_RNN}. It is defined as $W = 4n_{c}^2 + 4 n_i n_c + n_c n_o + 3 n_c $, where
$n_c$, $n_i$, and $n_o$ denote the number of memory cells, the number of input units, and the number of output units, respectively. As we have taken one LSTM layer with 100 hidden neurons, 4 input features, and $K + 1$ output classes, $n_c = 100$, $n_i =4$, and $n_o=K+1$. The number of fixed input weights, recurrent weights, and peephole weights in $W$ are $4$, $4$, and $3$, respectively \cite{Greff_LSTM}. 
In the conventional search, we find the optimal transmitter among all candidates, directly using (\ref{OS_scheme}) with global CSI. The complexity of evaluation of the values $C_s^k$ and their comparison to get the maximum are both $\mathcal{O}(K)$, and therefore the complexity of this approach is $\mathcal{O}(K)$. 
In the case where a large number of transmitters is selected, the complexity of the conventional scheme increases significantly due to combinatorial search, and can be written as $\mathcal{O}(K + \binom{K}{n} \log \binom{K}{n})$, where $n$ is the number of selected transmitters out of $K$.




\begin{table}[]

\centering
 \caption{Complexities of different TS schemes.}
\begin{tabular}{ |c|c|}
 \hline
 Scheme  & Complexity\\ 
 \hline
  LSTM &  $\mathcal{O}(KW)$\\ 
  \hline
  DNN & $\mathcal{O}(NL_1 + L_1L_2 + L_2K)$\\
  \hline
  SVM & $\mathcal{O}(N^2)$ \\
  \hline
  $k$-NN & $\mathcal{O}(N)$ \\
  \hline
  NB & $\mathcal{O} ((K+1)N + K)$\\
  \hline
  Conventional & $\mathcal{O} (K)$\\
 \hline
 \end{tabular}
\vspace{-.5cm}
 \label{table1}
\end{table}

\section{Results and Discussions}
This section compares LSTM, DNN, SVM, $k$-NN, and NB schemes with the conventional benchmark scheme. A training data set size of $M = 10^5$ is considered and the SOP is plotted by averaging the results from $M = 10^6$ batches of test data.
The parameters for the traditional ML (SVM, $k$-NN, and NB) and DNN schemes are the same as those in \cite{Yao_ML} and \cite{Yao_DNN}, respectively. 
For the LSTM parameters, activation functions for all gates are taken to be sigmoid, whereas the hyperbolic tangent is used for the memory cell in accordance with the standard LSTM architecture \cite{Sak_RNN}. Further, the Adam optimizer is implemented to find the optimum weights in the LSTM model \cite{BaekCSI_LSTM}. We have implemented one LSTM layer with $100$ hidden neurons and set the maximum number of epochs to $20$. The initial learning rate and minibatch size are kept identical with that of the DNN.

For the physical layer system parameters, unless otherwise specified, $R_{th} = 0.5$, and $\Phi = 0.1$. The channel parameters for the independent identically distributed (IID) channel case are  $(1/\lambda_{tr}, 1/\lambda_{td}, 1/\lambda_{sd}, 1/\lambda_{sr}, 1/\lambda_{se}, 1/\lambda_{te}) = (3, -6, 3, -3, -3, 6)$ dB, where $1/\lambda_{s_kd}=1/\lambda_{sd}$, $1/\lambda_{s_ke}=1/\lambda_{se}$, $1/\lambda_{s_kr}=1/\lambda_{sr}$, $\forall k$. 
In the INID channel case, the channel parameters are $(1/\lambda_{tr}, 1/\lambda_{td}, 1/\lambda_{sr}, 1/\lambda_{te}) = (3, -6, -3, 6)$  dB, $1/\lambda_{s_{k}d} = ( 0, 3, 6, 9)$ dB, $1/\lambda_{s_{k}e} = ( -6, -3, 0, 3)$ dB, $\forall k$, and $\delta = (0.90, 0.92, 0.94, 0.96)$ when $K=4$, and $1/\lambda_{s_{k}d} = ( -12, \ldots, 21)$ dB at a difference of 3 dB, $1/\lambda_{s_{k}e} = ( -21,  \ldots, 12)$ dB at a difference of 3 dB, $\forall k$, and $\delta = (0.78, \ldots, 1 )$ at a difference of 0.2 when $K = 12$.
\begin{figure*}[]
\label{fig_mis}
     \subfloat[IID channel case when $K$ = $4$ and $\delta$ = $0.8$.]{
     \begin{minipage}[l]{0.7\columnwidth}
         \centering
         \includegraphics[width=4cm, height=4cm ]{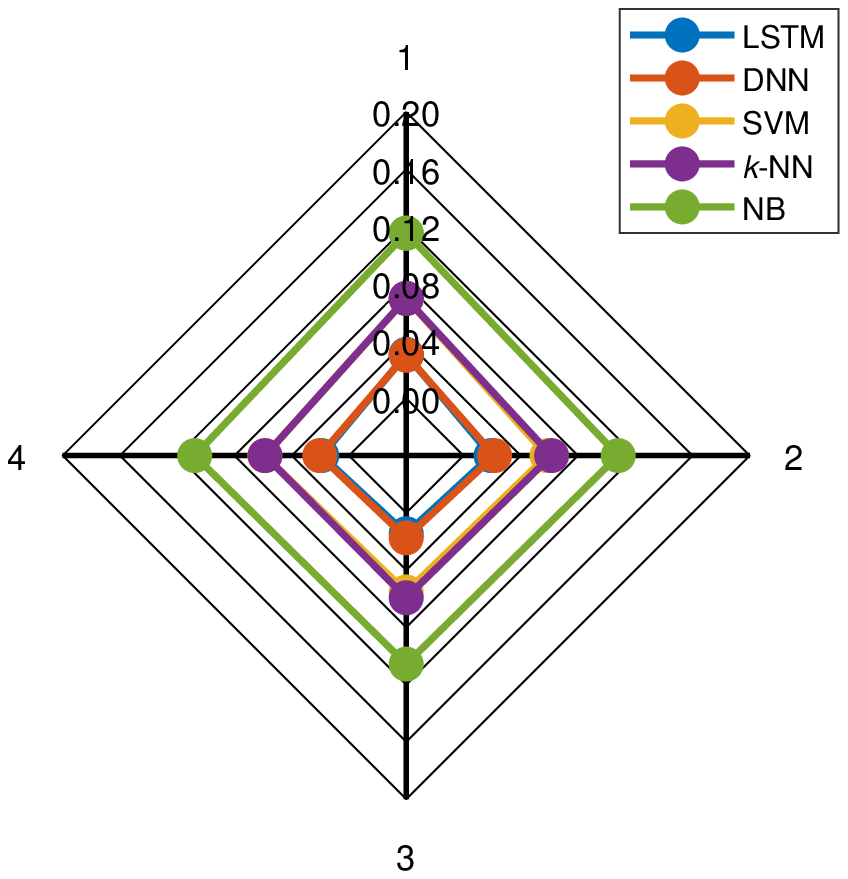}
          \label{fig:AAA}
     \end{minipage}}
     \subfloat[IID channel case when $K$ = $12$ and $\delta$ = $0.8$.]{
     \begin{minipage}[c]{0.7\columnwidth}
         \centering
         \includegraphics[width=4cm, height=4cm ]{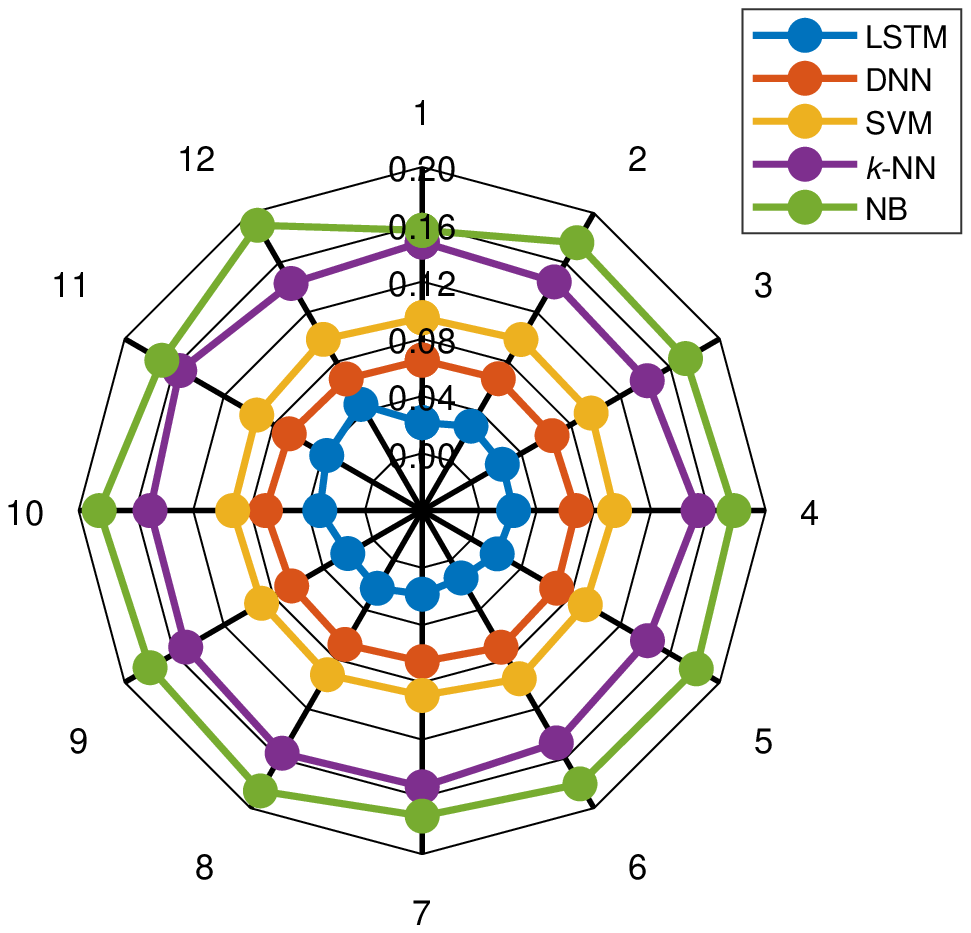}
        \label{fig:BBB}
     \end{minipage}}
     \subfloat[INID channel case when $K = 12$.]{
          \begin{minipage}[r]{0.7\columnwidth}
         \centering
         \includegraphics[width=4cm, height=4cm ]{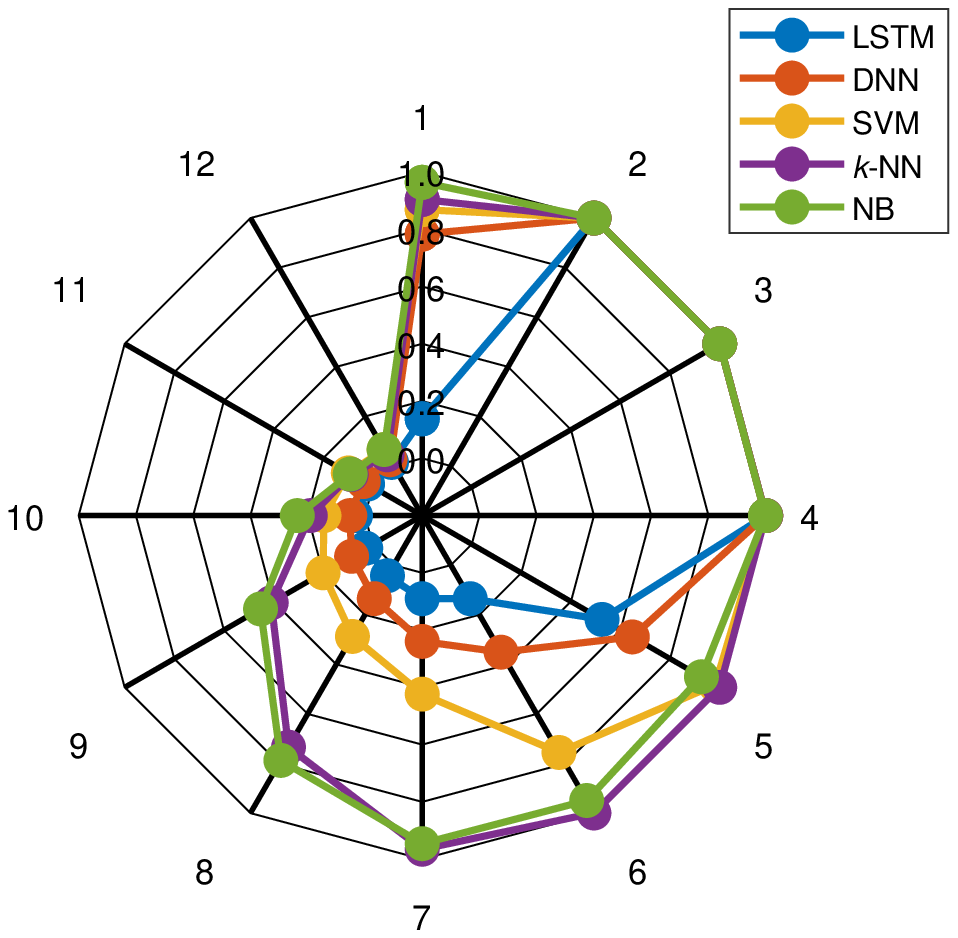}
        \label{fig:CCC}
     \end{minipage}}
      \vspace{-.1 cm}
     \caption{Misclassification rate for TS at $\Gamma_T=8$ dB.}
      \vspace{-.4cm}
\end{figure*}

 \begin{figure}
 \centering 
 \includegraphics[width=2.7in]{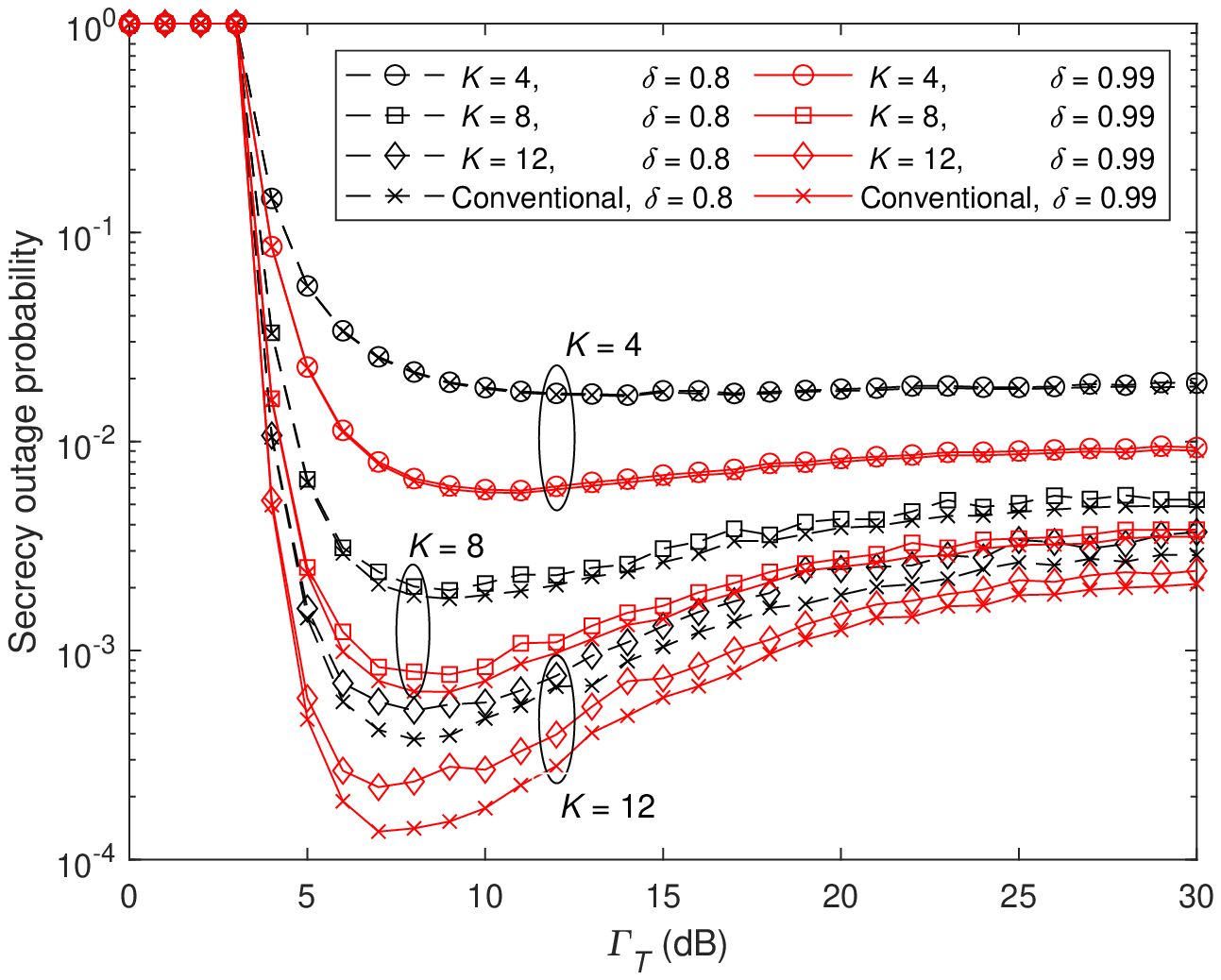} 
\vspace{-.2 cm}
 \caption{SOP versus $\Gamma_T$ performance of LSTM based model for different $K$ and $\delta$ in IID channel case.}
 \vspace{-.7cm}
 \label{fig_result1}
 \end{figure}

Figs. \ref{fig:AAA}-\ref{fig:CCC} present the misclassification rate of the learning schemes for $K$ transmitters at $\Gamma_T=8$ dB. Figs. \ref{fig:AAA} and \ref{fig:BBB}
are obtained for the IID channel case at $K=4$ and $12$, respectively, whereas Fig. \ref{fig:CCC} is for the INID channel case with $K=12$. Specifically, $\Gamma_T=8$ dB is considered as it provides minimum SOP at the chosen parameter values. In the figure, each vertex of the polygon represents the misclassification for each transmitter index.
The corresponding misclassification rate is presented on the straight line passing through the center of the polygon to the vertex.
The misclassification is considered when the learning model labels an input data wrongly, i.e., $\ell\rightarrow\bar{\ell}$, where $\forall\bar{\ell} \in \mathcal{L}$ such that $ \ell\ne \bar{\ell}$ and $\mathcal{L}$ is the set of all labels.

Comparing Figs. \ref{fig:AAA}-\ref{fig:CCC}, we can observe that the misclassification rate is more or less the same in each class in the IID channel case, but this is not true for the INID case. Due to the IID channel case, the number of CSI samples belonging to each class are more or less same; however, that is not true for the INID case. Hence, the misclassification rate is also different for INID.
However, it is clearly observed in the IID channel case that when the number of transmitters increases from $K=4$  to $K=12$, the misclassification rate increases. This leads to increased SOP performance degradation and will be explained during the discussion of Fig. \ref{fig_result1}.

Note from Fig. \ref{fig:BBB} that when the number of transmitters is large, i.e., $K=12$, LSTM performs the best as a classifier (it results in the polygon with the least area). We can also verify that in general, traditional ML based schemes perform worse than DL based schemes. Among the traditional ML schemes, the SVM model has achieved the
best performance, $k$-NN performs moderately, and NB
is the worst. Since
we have a finite number ($4K$) of training samples of random channel gains with
large dimensions, it is highly likely that samples may
contain some outliers. The SVM classifier has the capability
of handling such outliers and also can support high dimensional
data sets, hence; SVM performs best among the traditional ML schemes. However, the $k$-NN classifier often suffers
from high variations of classification when the dimension of the data is large, thus explaining its moderate performance \cite{Joung2016}.
Due to limited samples, the probabilities calculated
by the NB classifier are often inaccurate, and hence, returns worst
performance \cite{Yao_ML}.
DNN performs better than the traditional ML schemes
because of its use of multiple hidden layers, which
helps to extract the features for classification with
high accuracy. LSTM has memory units which result
in remembering the input information and identifying long-term dependency in the input data set \cite{Hochreiter}.  This property is successfully utilized in our proposed model, as the secrecy rates of individual transmitters are related due to the common interference term coming from the secondary transmitter. This data dependency cannot be exploited fully during classification in the traditional ML models or in DNN; therefore, LSTM outperforms the other learning schemes.

Fig. \ref{fig_result1} shows the SOP versus $\Gamma_T$ performance for the LTSM scheme with $K=4$, $K=8$, and $K=12$ for two different backhaul uncertainty values, $\delta=0.8$ and $\delta=0.99$ in the IID channel case. We can see that when $K=4$, the SOP obtained using the LSTM scheme matches with the SOP obtained from the conventional search at a particular backhaul uncertainty. However, as the number of transmitters increases, the LSTM scheme overestimate the SOP; moreover, the gap between these two results increases. This observation conforms with the  misclassification diagram of $K=4$ and $K=12$ in Fig. \ref{fig:AAA} and Fig. \ref{fig:BBB}, respectively. This is reasonable because as the number of possible classes increases, the probability of confusion between two classes also increases during a learning based decision making process. As a result, the SOP mismatch increases. In addition, we also observe for a particular $K$ that as the backhaul reliability decreases from $\delta=0.99$ to $0.8$, the gap between SOPs of LSTM and conventional search decreases. This shows that at a lower backhaul reliability, the misclassification rate decreases. At a low backhaul reliability, fewer transmitters are active, which reduces the data points in a particular feature vector and makes it sparse; as a consequence, LSTM can classify feature vectors accurately and can minimize decision error.

\begin{figure}
 \centering 
 \includegraphics[width=2.7in]{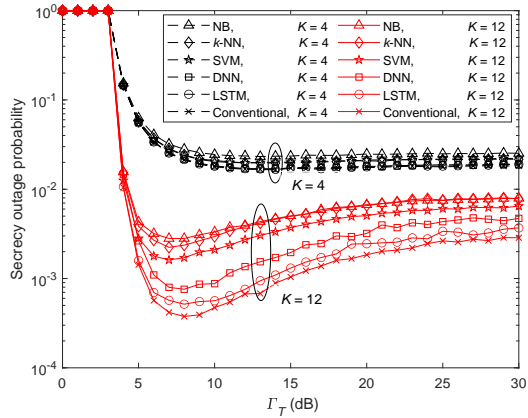} 
    \vspace{-.2 cm}
 \caption{SOP versus $\Gamma_T$ for different $K$ with $\delta = 0.8$ in IID channel case.}
 \vspace{-.4cm}
 \label{fig_result2}
 \end{figure}
 
Fig. \ref{fig_result2} compares SOP versus $\Gamma_T$ of the proposed learning schemes for $K=4$ and $K=12$ in the IID channel case. Here, as in Fig. \ref{fig_result1}, we can see that the SOP results of the learning schemes are closer to the conventional search result when the number of transmitters is less, and the gaps between them increase when the number of transmitters increases. This means that the misclassification rate in learning models is in general lower when a smaller number of transmitters (i.e., classes) exists, and that the misclassification rate increases with an increase in the number of transmitters. As expected, the LSTM technique performs the best, outperforming even the DNN scheme. This is because the LSTM scheme can exploit the correlated structure of the feature vectors better than the DNN scheme due to its internal recurrent structure. In general, it is verified that the performance of DL schemes, LSTM and DNN, are much better than the performance of the traditional ML techniques, SVM, $k$-NN, and NB, for TS.

  \begin{figure}
 \centering 
 \includegraphics[width=2.7in]{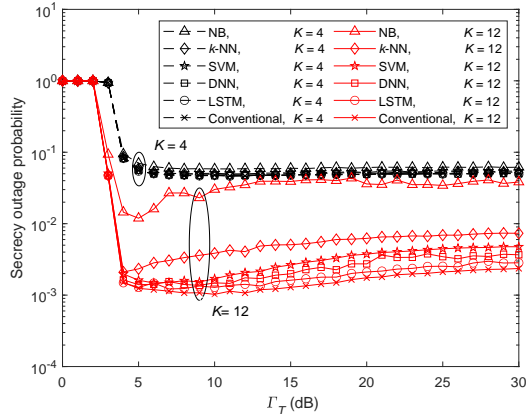} 
 \vspace{-.2 cm}
 \caption{SOP versus $\Gamma_T$ for different $K$ and $\delta = 0.8$ in INID channel case.}
   \vspace{-.6 cm}
 \label{fig_result4}
 \end{figure}
 
The SOP results of Fig. \ref{fig_result2} are replicated for the INID channel condition in Fig. \ref{fig_result4}. The observations are broadly the same as for Fig. \ref{fig_result2}. However, it can be observed that the NB technique performs worse in the INID channel conditions.



\section{Conclusion}
In this work, we have applied LSTM, an advanced version of conventional RNN based learning model, to a multiclass-classification problem of TS to improve security in a small-cell CR network with unreliable backhaul. Misclassification and secrecy outage performance of the model is compared with three traditional ML based schemes, NB, $k$-NN, and SVM, and a DL based scheme, DNN.  
We observe that LSTM matches the performance of the conventional technique when the number of transmitters is low.
Further, the performance of all ML schemes degrades as the number of transmitters increases as compared to the conventional scheme; however, LSTM has the best performance. We observe that DL based schemes perform better than traditional ML based schemes as they can handle nonlinear problems and large data set.  
We further notice that the LSTM based model significantly outperforms the DNN based model as the number of transmitters grows due to the memory based recurrent architecture. We conclude that the LSTM based TS is the best suited when a large number of classes with data dependency exist. It can also reduce feedback overhead against the conventional TS scheme.  
 \vspace{-.1cm}

\section*{Acknowledgement}
This publication has emanated from research conducted with the financial support of Science Foundation Ireland (SFI) under Grant Number 17/US/3445.



\bibliographystyle{IEEEtran}
\bibliography{IEEEabrv,COG_BACKHAUL}
}
\end{document}